\documentclass[12pt,titlepage]{article}
\usepackage{bm, amssymb,pifont,cancel, amsmath,comment,color}
\usepackage[dvips]{graphicx}
\makeatletter

\newcommand{\cD}{\mathcal{D}}

\@addtoreset{equation}{section}
\makeatother
 
\begin{document}

\begin{titlepage}
\null
\begin{flushright}
WU-HEP-15-07
\end{flushright}

\vskip 1.5cm
\begin{center}
\baselineskip 0.8cm
{\LARGE \bf
Impacts of supersymmetric higher derivative terms on inflation models in supergravity}

\lineskip .75em
\vskip 1.5cm

\normalsize

{\large Shuntaro Aoki} $\!${\def\thefootnote{\fnsymbol{footnote}}\footnote[1]{E-mail address: shun-soccer@akane.waseda.jp}}, 
{\large and} {\large Yusuke Yamada} $\!${\def\thefootnote{\fnsymbol{footnote}}\footnote[2]{E-mail address: yuusuke-yamada@asagi.waseda.jp}}

\vskip 1.0em

{\small\it Department of Physics, Waseda University, \\ 
Tokyo 169-8555, Japan}

\vspace{12mm}

{\bf Abstract}\\[5mm]
{\parbox{13cm}{\hspace{5mm} \small
%%%%%%%%%%%%%%%%%%%%%%%%%%%%%%%%%%%%%%%%%%%%%%%%%%%%%%%%%%%%%%%%%%%%
We show the effects of supersymmetric higher derivative terms on inflation models in supergravity. The results show that such terms generically modify the effective kinetic coefficient of the inflaton during inflation if the cut off scale of the higher derivative operators is sufficiently small. In such a case, the $\eta$-problem in supergravity does not occur, and we find that the effective potential of the inflaton generically becomes a power type potential with a power smaller than two.
%%%%%%%%%%%%%%%%%%%%%%%%%%%%%%%%%%%%%%%%%%%%%%%%%%%%%%%%%%%%%%%%%%%%
}}

\end{center}

\end{titlepage}

\tableofcontents
\vspace{35pt}
\hrule
\section{Introduction}
Cosmic inflation~\cite{Guth:1980zm} is the most plausible paradigm for the early universe, which solves the initial condition problems in the standard Big-Bang scenario. Especialy, the so-called slow-roll inflation models, first proposed in Ref.\cite{Linde:1983gd}, are favored by the recent results of cosmic microwave background (CMB) observations, because the models naturally predict the density fluctuation with a spectrum compatible with the data. 

From a phenomenological and theoretical viewpoints, the inflation models in supersymmetric (SUSY) theory have been intriguingly studied so far. Supergravity (SUGRA) naturally appears as an effective theory of superstring, which is a possible quantum gravity theory, and therefore, the inflation in superstring theory may be described by SUGRA models.

In SUGRA, many non-renormalizable interactions appear due to the SUGRA effects, which play important roles in some cases. As non-renormalizable terms, terms including derivative operators more than two are also possible if ghost modes do not appear. Such a ghost free SUSY higher derivative term of chiral multiplets was found and studied in Refs.~\cite{Khoury:2011da,Farakos:2012je,Koehn:2012ar,Farakos:2012qu,Koehn:2012te,Farakos:2013fne}. Especially, it is known that the SUSY higher derivative terms studied in Ref.~\cite{Khoury:2011da} appear in the Dirac-Born-Infeld type action~\cite{Born:1934gh} of a chiral multiplet, which is expected to be an effective action of the D3-brane in superstring theory~\cite{Rocek:1997hi}. The effects of such terms on inflation have been studied in some works e.g.~ \cite{Koehn:2012np,Gwyn:2014wna,Aoki:2014pna}. 

Due to SUSY, the higher derivative extension also provides some non-trivial terms, which are not expected. In our previous work~\cite{Aoki:2014pna}, we focused on the fact that such SUSY higher derivative terms provide quadratic kinetic terms of chiral multiplets with non-minimal kinetic coefficients. Then, even if the K\"ahler potential does not have terms of a chiral multiplet, it can be dynamical, and therefore, we discussed the case that the K\"ahler potential does not include the inflaton multiplet. As a consequence of non-minimal kinetic coefficients, we found that the scalar potential become a nontrivial form in terms of the canonically normalized inflaton. We also found that the so-called $\eta$-problem caused by the factor $e^K$ in F-term potential  is absent in such a case.  

It is natural to consider whether such effects work in the presence of the K\"ahler potential including inflaton multiplet. In this work, we will discuss such an issue. As we will show below, the non-minimal kinetic term can be dominant when the cut off scale of the SUSY higher derivative is sufficiently small (but, larger than the Hubble scale during inflation). Especially, we focus on the models with two chiral multiplets denoted by $\Phi$ and $S$ with a superpotential $f(\Phi)S$ as in Refs.~\cite{Kawasaki:2000yn,Kallosh:2010xz}, in order to understand how the inflaton dynamics changes in the presence of such effects. 

We will find that, under the assumption that the cut off scale is sufficiently small, the model in Ref.~\cite{Aoki:2014pna} is effectively realized and therefore the usual $\eta$-problem is absent. Interestingly, except for the case that the inflation potential is dominated by a constant term, the effective potential of the canonically normalized inflaton $\varphi$ becomes $V\sim \varphi^{2m/(m+2)}$ ($m\geq1$). The value of $m$ is determined by the structure of the potential as we will discuss in Sec.\ref{universality}. 

The remaining parts of this paper are organized as follows. First, we will briefly review the ghost free SUSY higher derivative terms in Sec.~\ref{review}. As we will find, such terms provide higher order terms of the F-terms, which cause a difficulty of solving the equations of auxiliary fields. We summarized the calculation for solving the equations in Appendix.~\ref{F-term}. By using the result, we will show the on-shell action with SUSY higher derivative terms in Sec.\ref{effective} and find that non-minimal kinetic terms of chiral multiplets appear. Then, we will discuss the condition with which the non-minimal kinetic term sufficiently works. Under such a condition, we will find that the field value of the inflaton generically becomes smaller than the Planck scale. Taking into account such observations, we will investigate a property of the effective action in Sec.~\ref{universality}, and find that the canonically normalized inflaton potential can only take a form mentioned above. Finally we will summarize this paper in Sec.~\ref{summary}.
%%%%%%%%%%%%%%%%%%%%%%%%%%%%%%%%%%%%%%%%%%%%%%%%%%%%%%%%%%%%%%%%%%%%%%%%%%%%%%%%%%%%%%%%%%%%%%%%%%%%%%%%%%%%%%%%%%%%%%%%%%%%%%%%%%%
\section{Brief review of ghost free SUSY higher derivative terms}\label{review}
In this section, we review the ghost-free SUSY higher derivative terms studied in Refs.~\cite{Khoury:2011da,Koehn:2012ar,Koehn:2012np}.\footnote{For more detailed review, see also Refs.~\cite{Koehn:2014apa,Koehn:2014zoa}} Especially, we focus on the higher derivative term which is independent of the difference of the SUGRA formulations, such as the old-, new-, and non-minimal one.  To clarify the universality of the following discussion in different SUGRA formulations, we discuss the higher derivative term in conformal SUGRA~\cite{Kaku:1978nz,Kugo:1982cu}, in which the different formulations are understood as the conformal SUGRA with different compensator multiplets~\cite{Kugo:1982cu}. The ghost-free SUSY higher derivative term can be written as follows,
\begin{align}
[T_{IJ\bar{K}\bar{L}}(\Phi,\bar{\Phi})\cD^\alpha\Phi^I\cD_\alpha\Phi^J\bar{\cD}_{\dot{\alpha}}\bar{\Phi}^{\bar{K}}\bar{\cD}^{\dot{\alpha}}\bar{\Phi}^{\bar{L}}]_D,\label{cHDT}
\end{align}
where $[\cdots]_D$ denotes the D-term density formula~\cite{Kugo:1982mr}, $\Phi$ and $\bar{\Phi}$ denote the chiral and anti-chiral matter multiplets with the Weyl ($w=0$) and chiral weight ($n=0$) respectively, $T_{IJ\bar{K}\bar{L}}(\Phi,\bar{\Phi})$ is a real function of $\Phi$ and $\bar{\Phi}$, $I,J$ ($\bar{K},\bar{L}$) are indices of (anti-)chiral multiplet, and $\cD_\alpha$ is the superconformal spinor derivative with a spinor index $\alpha$ defined in Ref.~\cite{Kugo:1983mv}. The indices $I$ and $J$ (or $\bar{K}$ and $\bar{L}$) are symmetric in $T_{IJ\bar{K}\bar{L}}$. Note that SUGRA terms often couple to the compensator multiplets, however the term in Eq.~(\ref{cHDT}) does not contain a compensator multiplet. As shown in Ref.~\cite{Ferrara:1983dh}, the different SUGRA formulations are equivalent to each other if no derivative operators on compensator are contained. Therefore, the term in Eq.~(\ref{cHDT}) is not affected by the difference of the compensator, that is, the difference of the formulations.

Let us discuss the bosonic component action of the term in Eq.~(\ref{cHDT}) denoted by $S_{HD}$. It is given by 
\begin{align}
S_{HD}=\int d^4x \sqrt {-g}\left[ 32T_{IJ\bar{K}\bar{L}}(F^IF^J\bar{F}^{\bar{K}}\bar{F}^{\bar{L}}-2F^I\bar{F}^{\bar{K}}\partial_\mu\Phi^J\partial^\mu\bar{\Phi}^{\bar{L}}+\partial_\mu\Phi^I\partial^\mu\Phi^J\partial_\nu\bar{\Phi}^{\bar{K}}\partial^\nu\bar{\Phi}^{\bar{L}})\right] ,\label{HDTcomp}
\end{align}
where $g={\rm det}g_{\mu\nu}$ and $g_{\mu\nu}$ is the spacetime metric, $F^I$ denotes the F-term of the chiral multiplet $\Phi^I$ and we use same letter for a multiplet $\Phi$ and its scalar component. 

We emphasize that, in Eq~(\ref{HDTcomp}), $S_{HD}$ contains not only higher derivative terms ($\sim T_{IJ\bar{K}\bar{L}}\partial_\mu\Phi^I\partial^\mu\Phi^J\partial_\nu\bar{\Phi}^{\bar{K}}\partial^\nu\bar{\Phi}^{\bar{L}}$) but also the quadratic derivative terms ($\sim T_{IJ\bar{K}\bar{L}}F^I\bar{F}^{\bar{K}}\partial_\mu\Phi^J\partial^\mu\bar{\Phi}^{\bar{L}}$). The quadratic derivative terms are the most important part in the following discussion. 

We also note that there are quartic couplings of F-terms in Eq.~(\ref{HDTcomp}), and then the equation of motion (E.O.M) of the auxiliary field $F^I$ becomes a cubic equation. Therefore, the on-shell action is not uniquely determined unlike the case without any higher derivative couplings. In the following discussion, we will choose the solution of E.O.M of F-term not containing any singularities at the limit $T_{IJ\bar{K}\bar{L}}\rightarrow 0$.

For concreteness of the following discussion, we consider models in the old minimal SUGRA formulation, which corresponds to the conformal SUGRA models with a chiral compensator. The general action of chiral matter multiplets without any higher derivative terms is written as
\begin{align}
\left[ -\frac{3}{2}S_0\bar{S}_0e^{-K/3}\right]_D+[S_0^3W]_F,\label{gen}
\end{align}
where $S_0$ denotes the chiral compensator with $(w,n)=(1,1)$, $K$ is the real function of $\Phi$ and $\bar{\Phi}$ called K\"ahler potential, and $W$ is the superpotential which is a holomorphic function of $\Phi$. The bosonic part of the action~(\ref{gen}), denoted by $S_{m}$, is
\begin{align}
S_m=&\int d^4x \sqrt{-g}\Biggl[\frac{1}{2}S_0\bar{S}_0e^{-K/3}R+3e^{-K/3}(D_\mu S_0D^\mu\bar{S}_0-F^{S_0}\bar{F}^{S_0})\nonumber\\
&-S_0\bar{S}_0e^{-K/3}(K_{I\bar{J}}-K_IK_{\bar{J}}/3)(\partial_\mu\Phi^I\partial^\mu\bar{\Phi}^{\bar{J}}-F^I\bar{F}^{\bar{J}})\nonumber\\
&-\left(\bar{S}_0K_{\bar{J}}e^{-K/3}(D_\mu S_0D^\mu\bar{\Phi}^{\bar{J}}-F^{S_0}\bar{F}^{\bar{J}})+{\rm h.c.}\right)+\left(S_0^3W_IF^I+3S_0^2F^{S_0}W+{\rm h.c.}\right)\Biggr],\label{comp1}
\end{align}
where $R$ is the Ricci scalar, $D_\mu S_0=(\partial_\mu-b_\mu-iA_\mu)S_0$, $b_\mu$ and $A_\mu$ are gauge fields of the dilatation and U(1)$_A$ symmetry respectively, $K_I=\partial_IK$, $W_I=\partial_IW$, and $K_{I\bar{J}}=\partial_I\partial_{\bar{J}}K$. As the superconformal gauge fixing conditions, we choose the following conventional ones~\cite{Kugo:1982mr} to obtain the action in the Einstein frame,
\begin{align}
S_0=\bar{S}_0=e^{K/6},\ b_\mu=0.
\end{align}
After substituting the conditions, because of the absence of the compensator dependent terms in Eq.~(\ref{HDTcomp}), we can integrate out $A_\mu$, $F^{S_0}$, and its conjugate and obtain the partially on-shell action as follows,
\begin{align}
S_m=\int d^4x \sqrt{-g}\Biggl[\frac{1}{2}R-K_{I\bar{J}}(\partial_\mu\Phi^I\partial^\mu\bar{\Phi}^{\bar{J}}-F^I\bar{F}^{\bar{J}})+(e^{K/2}D_IWF^I+{\rm h.c.})+3e^K|W|^2\Biggr],
\end{align}
where $D_IW=W_I+K_IW$. Summing up $S_{HD}$ and $S_m$, we obtain the total bosonic action we will discuss as,
\begin{align}
S_{tot}=&\int d^4x \sqrt{-g}\Biggl[\frac{1}{2}R-K_{I\bar{J}}(\partial_\mu\Phi^I\partial^\mu\bar{\Phi}^{\bar{J}}-F^I\bar{F}^{\bar{J}})+(e^{K/2}D_IWF^I+{\rm h.c.})+3e^K|W|^2\nonumber\\
&+32T_{IJ\bar{K}\bar{L}}(F^IF^J\bar{F}^{\bar{K}}\bar{F}^{\bar{L}}-2F^I\bar{F}^{\bar{K}}\partial_\mu\Phi^J\partial^\mu\bar{\Phi}^{\bar{L}}+\partial_\mu\Phi^I\partial^\mu\Phi^J\partial_\nu\bar{\Phi}^{\bar{K}}\partial^\nu\bar{\Phi}^{\bar{L}})\Biggr].\label{comp2}
\end{align}
It is difficult to solve the E.O.Ms of F-terms in this generic action~(\ref{comp2}), and so we will discuss the on-shell action of a more specific model in Sec.\ref{effective} and Appendix \ref{F-term}.
%%%%%%%%%%%%%%%%%%%%%%%%%%%%%%%%%%%%%%%%%%%%%%%%%%%%%%%%%%%%%%%%%%%%%%%%%%%%%%%%%%%%%%%%%%%%%%%%%%%%%%%%%%%%%%%%%%%%%%%%%%%%%%%%%%%
\section{The effective kinetic coefficient during inflation}\label{effective}
In this section, we show that if the higher derivative term becomes sufficiently large, the running kinetic inflation~\cite{Takahashi:2010ky,Nakayama:2010kt} occurs generically with a particular kinetic coefficient.  As shown in the previous section, the SUSY higher derivative terms generically contain kinetic terms of scalars with non-minimal coefficients (see Eq.~(\ref{comp2})). The coefficients are proportional to the F-terms, and at least one of the F-terms should have a non-zero value during inflation. Therefore, the kinetic term of inflaton becomes non-minimal if the inflaton multiplet couples to the SUSY breaking multiplet in SUSY higher derivative terms.

To clarify our statement, we consider a model with two chiral multiplets $\Phi$ and $S$ with the following K\"ahler potential, for simplicity,
\begin{align}
K=K_1(\Phi,\bar{\Phi})+K_2(S,\bar{S}),\label{Kahler}
\end{align}
where $K_1$ and $K_2$ are real functions of $\Phi$ and $S$ respectively. Then the total system~(\ref{comp2}) becomes as follows,
\begin{align}
S_{tot}=&\int d^4x \sqrt{-g}\Biggl[\frac{1}{2}R-K_{\Phi\bar{\Phi}}(\partial_\mu\Phi\partial^\mu\bar{\Phi}-F^\Phi\bar{F}^{\bar{\Phi}})+(e^{K/2}D_\Phi WF^\Phi+{\rm h.c.})+3e^K|W|^2\nonumber\\
&\qquad \qquad\qquad-K_{S\bar{S}}(\partial_\mu S\partial^\mu\bar{S}-F^S\bar{F}^{\bar{S}})+(e^{K/2}D_S WF^S+{\rm h.c.})\nonumber\\
&\qquad\qquad\qquad+32T_{\Phi\Phi\bar{\Phi}\bar{\Phi}}(|F^\Phi|^4-2|F^\Phi|^2\partial_\mu\Phi\partial^\mu\bar{\Phi}+|\partial_\mu\Phi\partial^\mu\Phi|^2)\nonumber\\
&\qquad\qquad\qquad+32T_{SS\bar{S}\bar{S}}(|F^S|^4-2|F^S|^2\partial_\mu S\partial^\mu\bar{S}+|\partial_\mu S\partial^\mu S|^2)\nonumber\\
&\qquad\qquad\qquad+32T_{\Phi S\bar{\Phi}\bar{S}}(4|F^\Phi|^2 |F^S|^2-2|F^\Phi|^2\partial_\mu S\partial^\mu\bar{S}-2|F^S|^2\partial_\mu \Phi\partial^\mu\bar{\Phi}\nonumber\\
&\qquad\qquad\qquad-2F^\Phi\bar{F}^{\bar{S}}\partial_\mu S\partial^\mu\bar{\Phi}-2F^S\bar{F}^{\bar{\Phi}}\partial_\mu \Phi\partial^\mu\bar{S}-4\partial_\mu\Phi\partial^\mu S\partial_\nu\bar{\Phi}\partial^\nu\bar{S})\Biggr]
\end{align}
We assume that $|D_SW|\gg |D_\Phi W|$ during inflation, that is, $S$ plays the role of the SUSY breaking multiplet, and $S$ is stabilized during inflation, i.e. $|\partial_\mu S|^2\sim 0$. Under these assumptions, we can solve the E.O.Ms of $F^S$ and $F^\Phi$. We show the detailed analysis of solving the E.O.Ms in Appendix~\ref{F-term}.  

From Eqs.~(\ref{FS}) and (\ref{FP}), we obtain the following on-shell Lagrangian of the inflaton $\Phi$,
\begin{align}
\mathcal{L}_{\rm eff}=&-\left(K_{\Phi\bar{\Phi}}+64T_{\Phi S\bar{\Phi}\bar{S}}\frac{e^K|D_SW|^2}{B^2}\right)\partial_\mu\Phi\partial^\mu\bar{\Phi}+32T_{\Phi\Phi\bar{\Phi}\bar{\Phi}}|\partial_\mu\Phi\partial^\mu\Phi|^2\nonumber\\
&-\left(\frac{2}{B}-\frac{K_{S\bar{S}}}{B^2}\right)e^K|D_SW|^2+3e^K|W|^2+\frac{32 T_{SS\bar{S}\bar{S}}}{B^4}e^{2K}|D_SW|^4,\label{Leff}
\end{align}
where we have omitted the terms including $D_\Phi W$ and $\partial_\mu S$ because they are assumed to be small during inflation, and $B\equiv K_{S\bar{S}}-64T_{\Phi S\bar{\Phi}\bar{S}}|\partial_\mu\Phi|^2$.

In the following discussions, we focus on the slow-roll inflation models, and then the effective Lagrangian~(\ref{Leff}) of the inflaton $\Phi$ is approximately expressed as,
\begin{align}
\mathcal{L}_{\rm eff}\sim&-\left(K_{\Phi\bar{\Phi}}+64T_{\Phi S\bar{\Phi}\bar{S}}K_{S\bar{S}}^{-2}e^K|D_SW|^2\right)\partial_\mu\Phi\partial^\mu\bar{\Phi}\nonumber\\
&-e^KK_{S\bar{S}}^{-1}|D_SW|^2+3e^K|W|^2+\frac{32 T_{SS\bar{S}\bar{S}}}{K_{S\bar{S}}^4}|D_SW|^4,\label{Leff2}
\end{align}
where we have used the approximation $B\sim K_{S\bar{S}}$ and neglected the last term on the first line in Eq.~(\ref{Leff}). Let us focus on the kinetic term of the inflaton in Eq.~(\ref{Leff2}) given by,
\begin{align}
\mathcal{L}_{\rm kin}= -\left(K_{\Phi\bar{\Phi}}+64T_{\Phi S\bar{\Phi}\bar{S}}K_{S\bar{S}}^{-2}e^K|D_SW|^2\right)\partial_\mu\Phi\partial^\mu\bar{\Phi}.
\end{align}
As we discussed, the kinetic coefficient of the inflaton includes the contribution from the SUSY higher derivative terms, which depends on a part of the scalar potential~$e^KK_{S\bar{S}}^{-1}|D_SW|^2$. In the case that $T_{\Phi S\bar{\Phi}\bar{S}}\sim \mathcal{O}(1)$, the correction is negligibly small when $e^{K/2}|D_SW|\ll1$. However, this statement is not correct in the case of $T_{\Phi S\bar{\Phi}\bar{S}}\gg1$, which means the cut-off scale of the higher derivative correction is sufficiently smaller than the Planck scale $M_{pl}=1$. Let us consider the case in which $T_{IJ\bar{K}\bar{L}}=c_{IJ\bar{K}\bar{L}}/M^4$ where $c_{IJ\bar{K}\bar{L}}$ is a positive constant and $M$ is the cut-off scale. 

Let us quantify how small $M$ is required to affect the kinetic coefficient of the inflaton. The condition $M$ should satisfy is,
\begin{align}
K_{\Phi\bar{\Phi}}<64c_{\Phi S\bar{\Phi}\bar{S}}K_{S\bar{S}}^{-2}e^K|D_SW|^2/M^4,
\end{align}
equivalently,
\begin{align}
M<\left(\frac{64c_{\Phi S\bar{\Phi}\bar{S}}}{K_{\Phi\bar{\Phi}}K_{S\bar{S}}^2}\right)^{1/4}\left(e^K|D_SW|^2\right)^{1/4}.\label{Mcond}
\end{align}
During inflation, $V\sim e^K|D_SW|^2\sim H^2$ where $H$ is the Hubble parameter, and then the condition~(\ref{Mcond}) reads $M<\sqrt{H}$. On the other hand, $M$ should be larger than the Hubble scale $H$ for consistency of the effective theory. Therefore, for $M$ satisfying $H<M<\sqrt{H}$, the correction originated from SUSY higher derivative term affects the kinetic term of the inflaton. Especially, for $H<M\ll \sqrt{H}$, the kinetic coefficient is dominated by the correction $64c_{\Phi S\bar{\Phi}\bar{S}}K_{S\bar{S}}^{-2}e^K|D_SW|^2/M^4$. Note that the last term in Eq.~(\ref{Leff2}) gives negative contribution to the scalar potential and therefore we have to require the following condition to realize a successful inflation,
\begin{align}
 c_{SS\bar{S}\bar{S}}<\frac{M^4K_{S\bar{S}}^{2}}{32e^{K}K_{S\bar{S}}^{-1}|D_SW|^2}\sim \left(\frac{M}{\sqrt{H}}\right)^4\ll 1,
\end{align} 
and so we assume $c_{SS\bar{S}\bar{S}}=0$ in the following.

Under those assumptions, we finally obtain the following action,
\begin{align}
\mathcal{L}_{\rm eff}\sim& -\frac{64c_{\Phi S\bar{\Phi}\bar{S}}K_{S\bar{S}}^{-2}e^K|D_SW|^2}{M^4}\partial_\mu\Phi\partial^\mu\bar{\Phi}-e^KK_{S\bar{S}}^{-1}|D_SW|^2+3e^K|W|^2.\label{Leff3}
\end{align}
We find that the effective action is very similar to the one in a model suggested in our previous work~\cite{Aoki:2014pna}. In our previous model, we have assumed the absence of the K\"ahler potential term of the inflaton $\Phi$, that is, $K_1(\Phi,\bar{\Phi})=0$ in Eq.~(\ref{Kahler}) and the Volkov-Akulov property $S^2=0$ for $S$~\cite{Volkov:1973ix,Antoniadis:2014oya}. Surprisingly enough, in the present case, we do not require such conditions, however, the model in Ref.~\cite{Aoki:2014pna} is effectively realized. We note that the kinetic term of the inflaton becomes a standard one $-K_{\Phi\bar{\Phi}}\partial_\mu \Phi \partial^\mu \bar{\Phi}$ after the inflation ends and $V\sim e^K|D_SW|^2$ becomes small, unlike the case of Ref.~\cite{Aoki:2014pna}. The property is the same with the one of the running kinetic inflation~\cite{Takahashi:2010ky,Nakayama:2010kt}.

Let us discuss a generic property of this inflation model. Due to the non-minimal kinetic coefficient, the effective potential in terms of the canonical normalized scalar seems to be different from the one in terms of the original scalar field $\Phi$ and $\bar{\Phi}$. The kinetic term of the inflaton can be rewritten as 
\begin{align}
\mathcal{L}_{\rm kin}=&-\frac{64c_{\Phi S\bar{\Phi}\bar{S}}K_{S\bar{S}}^{-1}V}{M^4}\partial_\mu\Phi\partial^\mu\bar{\Phi}\nonumber\\
=&-\frac{192c_{\Phi S\bar{\Phi}\bar{S}}K_{S\bar{S}}^{-1}H^2}{M^4}\partial_\mu\Phi\partial^\mu\bar{\Phi},
\end{align} 
where we have used $e^{K}K_{S\bar{S}}^{-1}|D_SW|^2\sim V=3H^2$.  The canonical normalized scalar field $\tilde{\Phi}$ is related with $\Phi$ as
\begin{align}
d\tilde{\Phi}=\sqrt{192c_{\Phi S\bar{\Phi}\bar{S}}K_{S\bar{S}}^{-1}}\frac{H}{M^2}d\Phi.\label{relation}
\end{align}
We assumed $M^2\ll H$ to realize the effective running kinetic term, and then, the large field variation of $\tilde{\Phi}$ can be realized even with the small field variation of $\Phi$ due to the enhancement factor~$H/M^2$. For example, with $H\sim 10^{-4}$ and $M\sim 10^{-3}$, the enhancement factor becomes $H/M^2\sim 10^2$. To realize the large variation $|\Delta \tilde{\Phi}|\sim \mathcal{O}(10)$, the one of the original field $|\Delta\Phi|$ is $\mathcal{O}(0.1)$. In such a case, the $\eta$-problem caused by the factor $e^K$ does not occur because of the absence of the large field variation of the original field $\Phi$. This is a remarkable feature of our model.%%%%%%%%%%%%%%%%%%%%%%%%%%%%%%%%%%%%%%%%%%%%%%%%%%%%%%%%%%%%%%%%%%%%%%%%%%%%%%%%%%%%%%%%%%%%%%%%%%%%%%%%%%%%%%%%%%%%%%%%%%%%%%%%%%%
\section{Universal properties}\label{universality}
In this section, we show a concrete example to describe the mechanism we have discussed in Sec.~\ref{effective}. Here, let us consider the model with the following K\"ahler and super-potential terms,
\begin{align}
K=&\hat{K}(\Phi,\bar{\Phi})+|S|^2-\zeta|S|^4,\label{Kmodel}\\
W=&f(\Phi)S,\label{Wmodel}
\end{align}
where $\zeta$ is a real constant, and $f(\Phi)$ is a holomorphic function of $\Phi$. Due to the quartic coupling of $S$ in (\ref{Kmodel}), $S$ obtains a mass heavier than the Hubble scale during inflation as in the case of Refs.~\cite{Kawasaki:2000yn,Kallosh:2010xz}. Then, $S$ is stabilized at $S=0$, and we obtain the following effective action~(\ref{Leff3}),
\begin{align}
 \mathcal{L}_{\rm eff}\sim& -\frac{64c_{\Phi S\bar{\Phi}\bar{S}}e^{\hat{K}}|f(\Phi)|^2}{M^4}\partial_\mu\Phi\partial^\mu\bar{\Phi}-e^{\hat{K}}|f(\Phi)|^2,\label{model1}
 \end{align}
 where we have used $D_\Phi W=W=0$. 
 
 As discussed below Eq.~(\ref{relation}), even with the small field variation of $\Phi$, the variation of the canonically normalized scalar $\tilde{\Phi}$ can be much larger than $\mathcal{O}(1)$. In the following discussion, we assume that inflation ends at $\Phi\ll 1$. Under the assumption, to investigate the stage of inflation related to the cosmological parameters, we can expand the scalar potential with respect to $\Phi$ and $\bar{\Phi}$.\footnote{Even in the case that $\Phi|_{\rm end}\geq 1$, where $\Phi|_{\rm end}$ is the value of $\Phi$ at the end of inflation, we can perform a similar analysis by a field redefinition $\Phi-\Phi|_{\rm end}\to \Phi$.}
 Then, we obtain
 \begin{align}
 V=& e^{\hat{K}(0)}\left(|f(0)|^2+ \alpha\Phi+\bar{ \alpha} \bar{\Phi}+ \beta |\Phi|^2+\gamma \Phi^2+\bar{\gamma}\bar{\Phi}^2\right)+\mathcal{O}(\Phi^3)\nonumber\\
 \sim&e^{\hat{K}(0)}\Biggl[|f(0)|^2+2({\rm Re} \alpha)\phi-2({\rm Im} \alpha)\chi+( \beta+2{\rm Re}\gamma)\phi^2-4({\rm Im}\gamma)\phi\chi+( \beta-2{\rm Re}\gamma)\chi^2\Biggr]\nonumber\\
 \label{Veff}
 \end{align}
 where
 \begin{align}
 \alpha\equiv& \bar{f}(0)f'(0)+|f(0)|^2\hat{K}_\Phi(0),\\
 \beta\equiv&|f'(0)|^2+(\hat{K}_{\Phi\bar{\Phi}}(0)+|\hat{K}_\Phi(0)|^2)|f(0)|^2+(\bar{f}(0)f'(0)\hat{K}_{\Phi}(0)+{\rm h.c.}),\\
 \gamma\equiv&\frac{1}{2}\bar{f}(0)f''(0)+\frac{1}{2}|f(0)|^2(\hat{K}_{\Phi\Phi}+\hat{K}_\Phi^2)+\bar{f}(0)f'(0)\hat{K}_\Phi,
 \end{align}
 and we have defined $\phi$ and $\chi$ as $\Phi=\phi+i\chi$. For simplicity, we assume ${\rm Im}\alpha={\rm Im} \gamma=0$ which can be realized when the parameters in super- and K\"ahler potential are real. Then, the scalar potential (\ref{Veff}) becomes
 \begin{align}
 V\sim e^{\hat{K}(0)}\Biggl[|f(0)|^2+2({\rm Re}\alpha)\phi+( \beta+2{\rm Re}\gamma)\phi^2+( \beta-2{\rm Re}\gamma)\chi^2\Biggr].
 \end{align} 
 We identify $\phi$ as the inflaton, and then the effective action of the inflaton at $\chi=0$ is 
 \begin{align}
 \mathcal{L}_{\rm inf}\sim&  -e^{\hat{K}(0)}\Biggl[|f(0)|^2+2({\rm Re}\alpha)\phi+(\beta+2{\rm Re}\gamma)\phi^2\Biggr]\Biggl[\frac{64c_{\Phi S\bar{\Phi}\bar{S}}}{M^4}\partial_\mu\phi\partial^\mu\phi+1\Biggr].\nonumber\\
 \label{effaction}
 \end{align}

More general expression~(\ref{effaction}) is given by
\begin{align}
 \mathcal{L}_{\rm inf}\sim&  -\left[\sum_{n=0}^\infty V_n(0)\phi^n\right]\Biggl[\frac{64c_{\Phi S\bar{\Phi}\bar{S}}}{M^4}\partial_\mu\phi\partial^\mu\phi+1\Biggr].\nonumber\\
 \label{effaction2}
 \end{align}
where $V_n(0)=\partial^n_\phi V|_{\phi=0}/n!$. The canonically normalized inflaton $\varphi$ is determined by the dominant part of the scalar potential $V=\sum_{n=0}^\infty V_n(0)\phi^n$, and in the case that the dominant part is $V_m(0)\phi^m$, $\varphi$ is effectively expressed as
\begin{align}
\varphi\sim \int d\phi \frac{8\sqrt{c_{\Phi S\bar{\Phi}\bar{S}}V_m(0)}}{M^2}\phi^{\frac{m}{2}}=\frac{16\sqrt{c_{\Phi S\bar{\Phi}\bar{S}}V_m(0)}}{(m+2)M^2}\phi^{\frac{m}{2}+1}.
\end{align}
In the case that the dominant part of the inflaton potential is given by $V_0(0)$, $\varphi\sim C\phi$ where $C$ is a constant, then the scalar potential becomes $V(\phi)\sim V(\varphi/C)$ and the functional form is not changed. However, for $m\neq0$, the scalar potential is effectively given by
\begin{align}
V\sim V_m(0)\left(\frac{(m+2)M^2}{16\sqrt{c_{\Phi S\bar{\Phi}\bar{S}}V_m(0)}}\right)^{\frac{2m}{m+2}}\varphi^{\frac{2m}{m+2}}.\label{Veff2}
\end{align}
Surprisingly enough, the effective potential~(\ref{Veff2}) is one in the simplest chaotic inflation~\cite{Linde:1983gd} with the power $2/3\leq 2m/(m+2)<2$. 

As an example, we consider the model with
\begin{align}
\hat{K}&=|\Phi|^2\label{Km}\\
f(\Phi)&=\lambda\Phi\label{Wm} 
\end{align} 
where $\lambda$ is a real parameter. Note that the K\"ahler potential does not have the shift symmetry, which is required in the F-term inflation models~\cite{Kawasaki:2000yn,Kallosh:2010xz}. However, we can realize the inflation as discussed below. The action of the inflaton $\phi$ at $S=\chi=0$ becomes
\begin{align}
\mathcal{L}=-\left(1+\frac{64c_{\Phi S\bar{\Phi}\bar{S}}\lambda^2e^{\phi^2}\phi^2}{M^4}\right)\partial_\mu\phi\partial^\mu\phi-e^{\phi^2}\lambda^2\phi^2.\label{exact}
\end{align}
In this case, the leading order of the inflaton potential is
 \begin{align}
 V=\lambda^2e^{\phi^2}\phi^2\sim\lambda^2\phi^2
 \end{align}
 for $\phi\ll1$, which corresponds to the case with $m=2$ in Eq.~(\ref{Veff2}). Therefore the effective potential~(\ref{Veff2}) is expected to be a linear potential $\varphi$. 
Choosing the parameters as
\begin{align}
c_{\Phi S\bar{\Phi}\bar{S}}=1,\quad M=10^{-3}, \quad \lambda=1.09\times 10^{-3},
\end{align} 
we numerically solved the equation of motion for $\phi$, derived from the Lagrangian~(\ref{exact}), and confirmed that the model with K\"ahler potential (\ref{Km}) can realize the slow-roll inflation even without the shift symmetry for $\phi$. We obtained the following values of the spectral index $n_s$ and the tensor to scalar ratio $r$,
\begin{align}
n_s&=0.975\nonumber\\
r&=0.0665
\end{align}
where these values are ones at $N=60$ and $N$ denotes the number of e-foldings. The values of $n_s$ and $r$ are the same with that of a linear potential respectively, as we expected. We also obtained the value of $\phi$ at $N=60$ as
\begin{align}
\phi|_{N=60}=0.000596.
\end{align}
The smallness of $\phi|_{N=60}$ shows the validity of our discussion with the expansion in Eq.~(\ref{effaction}) and more generally Eq.~(\ref{effaction2}). The order of the value $\phi|_{N=60}$ seems to be too small to realize the large field inflation, however, the field variation of $\varphi$ is larger than 1 due to the enhancement factor $H/M^2\gg 1$. The ratio between Hubble scale and $M$ in this case is $M/H=37.6$, and therefore the cut off scale is much larger than the Hubble scale. 

Another property of our models is the existence of at least one light scalar field. We assumed that $\chi$ which is a imaginary part of $\Phi$. In a broad class of the SUGRA inflation models, scalar fields are stabilized during inflation by the Hubble induced mass $m\sim H$ from the factor $e^K$ in the F-term potential. However, due to the enhanced kinetic coefficient of $\chi$, the mass of $\chi$ becomes smaller than $H$ during inflation. Then, the light mode $\chi$ also quantum mechanically fluctuates, which leads to an isocurvature perturbation. In the case that $\chi$ dominates the energy density of the universe and decays to other light elements after reheating by $\phi$, the isocurvature mode can be an adiabatic one. Then such a curvature perturbation generically has the non-Gaussian property. It is interesting to discuss such a scenario, however it beyonds the scope of this paper. We will investigate such scenarios in future. 
%%%%%%%%%%%%%%%%%%%%%%%%%%%%%%%%%%%%%%%%%%%%%%%%%%%%%%%%%%%%%%%%%%%%%%%%%%%%%%%%%%%%%%%%%%%%%%%%%%%%%%%%%%%%%%%%%%%%%%%%%%%%%%%%%%%
\section{Summary}\label{summary}
We have investigated the effects of SUSY higher derivative terms on inflation models. Especially, we have focused on the models with two chiral multiplets, in which the K\"ahler and superpotential are given in Eqs.~(\ref{Kmodel}) and (\ref{Wmodel}). 

Our assumptions are summarized as follows. The action contains ghost free higher derivative terms~(\ref{cHDT}) and the coefficients are $T_{IJ\bar{K}\bar{L}}=c_{IJ\bar{K}\bar{L}}/M^4$, where the cut off scale $M$ satisfies the condition $H<M\ll \sqrt{H}$. To avoid the negative energy during inflation, a coupling constant $c_{SS\bar{S}\bar{S}}$ in Eq.(\ref{Leff2}) is set to be zero or sufficiently small. 

Under those assumptions, the non-minimal kinetic coefficient, induced by SUSY higher derivative terms, becomes the dominant part of the coefficients as shown in Sec.~\ref{effective}. Then the kinetic term of the inflaton is effectively expressed as $-c_{\Phi S\bar{\Phi}\bar{S}}(V/M^4)|\partial_\mu \Phi|^2$ where $V=3H^2$ is the scalar potential and $H^2/M^4\gg 1$. As in the case of the running kinetic inflation~\cite{Takahashi:2010ky,Nakayama:2010kt}, the kinetic term of the inflaton multiplet becomes a standard one after the inflation, however, the non-minimal kinetic term drastically changes the inflaton dynamics during inflation. In Sec.~\ref{effective}, we have found that even if the canonical normalized scalar $\tilde{\Phi}$ takes a field value larger than the Planck scale, the field value of $\Phi$ can be smaller than the Planck scale. Therefore, the $\eta$ problem associated with the F-term potential is absent even without shift symmetry of the inflaton, and one can expand the scalar potential with respect to $\Phi$ and find the expression given in Eq.~(\ref{Veff}). Except for the case that a constant term is the dominant part during inflation, the effective potential only can takes a form $V\sim \varphi^{2m/(m+2)}$ where $\varphi$ is the canonically normalized inflaton and $m\geq1$. This surprising and predictive result is almost independent of the function $f(\Phi)$, and is originated from the structure of the SUSY higher derivative terms. The potential $V\sim \varphi^{2m/(m+2)}$ with a small $m$ predicts the cosmological parameters compatible with the latest Planck results~\cite{Ade:2015lrj}, and will be tested by future CMB observations.
%%%%%%%%%%%%%%%%%%%%%%%%%%%%%%%%%%%%%%%%%%%%%%%%%%%%%%%%%%%%%%%%%%%%%%%%%%%%%%%%%%%%%%%%%%%%%%%%%%%%%%%%%%%%%%%%%%%%%%%%%%%%%%%%%%%
\section*{Acknowledgment}
YY would like to thank Hideo Kodama, Yoske Sumitomo, and Fuminobu Takahashi for stimulating discussion and comments at KEK-CPWS The 4th UTQuest Workshop "B-mode Cosmology". The work of YY was supported by JSPS
Research Fellowships for Young Scientists No. 26-4236 in Japan.
%%%%%%%%%%%%%%%%%%%%%%%%%%%%%%%%%%%%%%%%%%%%%%%%%%%%%%%%%%%%%%%%%%%%%%%%%%%%%%%%%%%%%%%%%%%%%%%%%%%%%%%%%%%%%%%%%%%%%%%%%%%%%%%%%%%
\begin{appendix}
\section{Integration of the F-terms}\label{F-term}
We show the derivation of the on-shell action including the higher order terms of F-terms in models with two chiral multiplets $\Phi$ and $S$ discussed in Sec.~\ref{effective}. As shown in Eq.~(\ref{comp2}), the SUSY higher derivative action includes the quartic couplings of F-terms. The E. O. M of $F^{\Phi,S}$ are given by
\begin{align}
&(K_{\Phi\bar{\Phi}}-64T_{\Phi\Phi\bar{\Phi}\bar{\Phi}}|\partial_\mu \Phi|^2+128T_{\Phi S\bar{\Phi}\bar{S}}|F^S|^2-64T_{\Phi S \bar{\Phi}\bar{S}}|\partial_\mu S|^2+64T_{\Phi\Phi\bar{\Phi}\bar{\Phi}}|F^\Phi|^2)F^\Phi\nonumber\\
&+e^{K/2}D_{\bar{\Phi}}\bar{W}-64T_{\Phi S\bar{\Phi}\bar{S}}F^S\partial_\mu\Phi\partial^\mu\bar{S}=0,\label{Feq1}\\
&(K_{S\bar{S}}-64T_{SS\bar{S}\bar{S}}|\partial_\mu S|^2+128T_{\Phi S\bar{\Phi}\bar{S}}|F^\Phi|^2-64T_{\Phi S \bar{\Phi}\bar{S}}|\partial_\mu \Phi|^2+64T_{S S\bar{S}\bar{S}}|F^S|^2)F^S\nonumber\\
&+e^{K/2}D_{\bar{S}}\bar{W}-64T_{\Phi S\bar{\Phi}\bar{S}}F^\Phi\partial_\mu S\partial^\mu\bar{\Phi}=0.\label{Feq2}
\end{align}
The equations are complicated and difficult to solve them, therefore we assume the following conditions mentioned in Sec.~\ref{effective},
\begin{align}
&|D_{S}W|\gg|D_\Phi W|,\label{assume}\\
&|\partial_\mu S|^2\sim 0.
\end{align}
Then, Eqs.~(\ref{Feq1}) and (\ref{Feq2}) are reduced as
\begin{align}
&(K_{\Phi\bar{\Phi}}-64T_{\Phi\Phi\bar{\Phi}\bar{\Phi}}|\partial_\mu \Phi|^2+128T_{\Phi S\bar{\Phi}\bar{S}}|F^S|^2+64T_{\Phi\Phi\bar{\Phi}\bar{\Phi}}|F^\Phi|^2)F^\Phi+e^{K/2}D_{\bar{\Phi}}\bar{W}=0,\label{Feq3}\\
&(K_{S\bar{S}}+128T_{\Phi S\bar{\Phi}\bar{S}}|F^\Phi|^2-64T_{\Phi S \bar{\Phi}\bar{S}}|\partial_\mu \Phi|^2+64T_{S S\bar{S}\bar{S}}|F^S|^2)F^S+e^{K/2}D_{\bar{S}}\bar{W}=0.\label{Feq4}
\end{align}
By multiplying $D_\Phi W$ and $D_SW$ to Eqs.~(\ref{Feq3}) and (\ref{Feq4}) respectively, we obtain the following equations,
\begin{align}
&(K_{\Phi\bar{\Phi}}-64T_{\Phi\Phi\bar{\Phi}\bar{\Phi}}|\partial_\mu \Phi|^2+128T_{\Phi S\bar{\Phi}\bar{S}}|F^S|^2+64T_{\Phi\Phi\bar{\Phi}\bar{\Phi}}|F^\Phi|^2)X+e^{K/2}|D_{\bar{\Phi}}\bar{W}|^2=0,\label{Feq5}\\
&(K_{S\bar{S}}+128T_{\Phi S\bar{\Phi}\bar{S}}|F^\Phi|^2-64T_{\Phi S \bar{\Phi}\bar{S}}|\partial_\mu \Phi|^2+64T_{S S\bar{S}\bar{S}}|F^S|^2)Y+e^{K/2}|D_{\bar{S}}\bar{W}|^2=0,\label{Feq6}
\end{align}
where $X=F^\Phi D_\Phi W$ and $Y=F^SD_SW$. In Eqs.~(\ref{Feq5}) and (\ref{Feq6}), all the quantity other than $X$ and $Y$ are real, and therefore $X$ and $Y$ should also be real. Then, we obtain the following equations with respect to $X$ and $Y$,
\begin{align}
&(K_{\Phi\bar{\Phi}}-64T_{\Phi\Phi\bar{\Phi}\bar{\Phi}}|\partial_\mu \Phi|^2+\frac{128T_{\Phi S\bar{\Phi}\bar{S}}}{|D_SW|^2}Y^2+\frac{64T_{\Phi\Phi\bar{\Phi}\bar{\Phi}}}{|D_\Phi W|^2}X^2)X+e^{K/2}|D_{\bar{\Phi}}\bar{W}|^2=0,\label{Feq7}\\
&(K_{S\bar{S}}+\frac{128T_{\Phi S\bar{\Phi}\bar{S}}}{|D_\Phi W|^2}X^2-64T_{\Phi S \bar{\Phi}\bar{S}}|\partial_\mu \Phi|^2+\frac{64T_{S S\bar{S}\bar{S}}}{|D_SW|^2}Y^2)Y+e^{K/2}|D_{\bar{S}}\bar{W}|^2=0.\label{Feq8}
\end{align}
We seek the solutions for $X$ and $Y$ which are continuously connected to the ones without SUSY higher derivative terms $F^\Phi=-e^{K/2}K^{\Phi\bar{\Phi}}D_{\bar{\Phi}}\bar{W}$, $F^S=-e^{K/2}K^{S\bar{S}}D_{\bar{S}}\bar{W}$. Therefore, we can estimate that $X$ and $Y$ are proportional to $|D_\Phi W |^2$ and $|D_SW|^2$. With this estimation and assumption~(\ref{assume}), we can neglect the second term in the parenthesis of Eq.~(\ref{Feq8}) and obtain the following equation closed in terms of $Y$,
\begin{align}
(K_{S\bar{S}}-64T_{\Phi S \bar{\Phi}\bar{S}}|\partial_\mu \Phi|^2+\frac{64T_{S S\bar{S}\bar{S}}}{|D_SW|^2}Y^2)Y+e^{K/2}|D_{\bar{S}}\bar{W}|^2=0.
\end{align}
We can easily solve this third order equation, and obtain 
\begin{align}
Y=&\sqrt{\frac{B}{3A}}\left(\sqrt{1+\frac{27AC}{4B^3}}-\frac{C}{2}\sqrt{\frac{27A}{B^3}}\right)^{\frac{1}{3}}\nonumber\\
&-\sqrt{\frac{B}{3A}}\left(\sqrt{1+\frac{27AC}{4B^3}}-\frac{C}{2}\sqrt{\frac{27A}{B^3}}\right)^{-\frac{1}{3}},\label{Ysol}
\end{align}
where $A=64T_{S S\bar{S}\bar{S}}/|D_SW|^2$, $B=K_{S\bar{S}}-64T_{\Phi S \bar{\Phi}\bar{S}}|\partial_\mu \Phi|^2$, and $C=e^{K/2}|D_{\bar{S}}\bar{W}|^2$. The solution~(\ref{Ysol}) has properties we required, that is, $Y\rightarrow -e^{K/2}K^{S\bar{S}}|D_SW|^2$ in the limit $T_{IJ\bar{K}\bar{L}}\rightarrow 0$. Indeed, in such a limit we can expand the expression~(\ref{Ysol}) as follows,
\begin{align}
Y\sim& \sqrt{\frac{B}{3A}}\left(1-\frac{C}{6}\sqrt{\frac{27A}{B^3}}\right)-\sqrt{\frac{B}{3A}}\left(1+\frac{C}{6}\sqrt{\frac{27A}{B^3}}\right)\nonumber\\
&=-\frac{C}{B}\nonumber\\
&\sim -e^{K}K_{S\bar{S}}^{-1}|D_SW|^2.
\end{align}
To make the discussion simple, we expand the solution~(\ref{Ysol}) with respect to $C$ in the same way discussed above, and obtain the approximated value of $Y$ as
\begin{align}
Y\sim -\frac{C}{B}=-\frac{e^{K/2}|D_SW|^2}{K_{S\bar{S}}-64T_{\Phi S\bar{\Phi}\bar{S}}|\partial_\mu \Phi|^2}.\label{Ysol2}
\end{align}
From Eq.~(\ref{Ysol2}), we obtain,
\begin{align}
F^S=-\frac{e^{K/2}D_{\bar{S}}\bar{W}}{K_{S\bar{S}}-64T_{\Phi S\bar{\Phi}\bar{S}}|\partial_\mu \Phi|^2}.\label{FS}
\end{align}
We can also solve Eq.~(\ref{Feq7}) with Eq.~(\ref{Ysol2}), and obtain the following solution,
\begin{align}
X=&\sqrt{\frac{\tilde{B}}{3\tilde{A}}}\left(\sqrt{1+\frac{27\tilde{A}\tilde{C}}{4\tilde{B}^3}}-\frac{\tilde{C}}{2}\sqrt{\frac{27\tilde{A}}{\tilde{B}^3}}\right)^{\frac{1}{3}}\nonumber\\
&-\sqrt{\frac{\tilde{B}}{3\tilde{A}}}\left(\sqrt{1+\frac{27\tilde{A}\tilde{C}}{4\tilde{B}^3}}-\frac{\tilde{C}}{2}\sqrt{\frac{27\tilde{A}}{\tilde{B}^3}}\right)^{-\frac{1}{3}},\label{Xsol}
\end{align}
where $\tilde{A}=64T_{\Phi\Phi\bar{\Phi}\bar{\Phi}}/|D_\Phi W|^2$, $\tilde{B}=K_{\Phi\bar{\Phi}}-64T_{\Phi \Phi \bar{\Phi}\bar{\Phi}}|\partial_\mu \Phi|^2+128T_{\Phi S\bar{\Phi}\bar{S}}Y^2/|D_SW|^2$, and $\tilde{C}=e^{K/2}|D_\Phi W|^2$. The solution~(\ref{Xsol}) is approximately given by
\begin{align}
X\sim-\frac{\tilde{C}}{\tilde{B}}=-\frac{e^{K/2}|D_\Phi W|^2}{K_{\Phi\bar{\Phi}}-64T_{\Phi \Phi \bar{\Phi}\bar{\Phi}}|\partial_\mu \Phi|^2+128T_{\Phi S\bar{\Phi}\bar{S}}Y^2/|D_SW|^2},\label{Xsol2}
\end{align}
and we obtain
\begin{align}
F^\Phi\sim-\frac{e^{K/2}D_{\bar{\Phi}} \bar{W}}{K_{\Phi\bar{\Phi}}-64T_{\Phi \Phi \bar{\Phi}\bar{\Phi}}|\partial_\mu \Phi|^2+128T_{\Phi S\bar{\Phi}\bar{S}}Y^2/|D_SW|^2}.\label{FP}
\end{align}
The solutions~(\ref{Ysol2}) and (\ref{Xsol2}) are consistent with our approximation, with which we neglect the second term in the parenthesis of Eq.~(\ref{Feq8}) under the assumption $|D_SW|\gg|D_\Phi W|$. 
\end{appendix}

\end{document}